\documentclass[onecolumn,amsmath,amssymb,superscriptaddress,nofootinbib,11pt,a4paper]{revtex4}
\usepackage{booktabs}
\usepackage{array}
\pdfoutput=1
\usepackage[T1]{fontenc}
\usepackage{CJK}
\usepackage{xcolor}
\usepackage{amsfonts}
\usepackage{epstopdf}
\usepackage{wrapfig} 
\usepackage{subcaption}
\usepackage{graphicx}  
\usepackage{dcolumn}   
\usepackage{bm}
\usepackage{float}
\usepackage[utf8]{inputenc}
\usepackage{orcidlink}

\begin{document}

\title{Extracting Energy from a Non-Kerr Rotating Spacetime with an Anomalous Quadrupole Moment via Magnetic Reconnection}

\author{Ke Wang}
\affiliation{School of Material Science and Engineering, Chongqing Jiaotong University, \\Chongqing 400074, China}

\author{Xiao-Xiong Zeng\footnote{Electronic address: xxzengphysics@163.com  (Corresponding author)}}
\affiliation{College of Physics and Optoelectronic Engineering, Chongqing Normal University, \\Chongqing 401331, China}

\begin{abstract}
{This paper investigates how to extract energy from a non-Kerr rotating spacetime with an anomalous quadrupole moment via the magnetic reconnection mechanism. Unlike many other rotating spacetimes, this spacetime possesses closed timelike curves, and the corresponding spacetime regions must be excluded when extracting energy. After introducing the event horizon, ergosphere, and closed timelike curves of this spacetime, we deeply analyze the energy per unit enthalpy at infinity for accelerated and decelerated plasmas, the allowed region for energy extraction, and the power and efficiency of energy extraction. The results show that energy extraction is possible for both positive and negative anomalous quadrupole moments, but a positive and small anomalous quadrupole moment corresponds to higher power and efficiency of energy extraction.
}
\end{abstract}

\maketitle
 \newpage
\section{Introduction}
Black holes are celestial objects predicted by Einstein's theory of gravity. A series of recent observations, such as the detection of gravitational waves by the Laser Interferometer Gravitational-Wave Observatory \cite{23,24} and the imaging of supermassive black holes by the Event Horizon Telescope \cite{25,26}, have confirmed the existence of black holes. Black holes play a central role in many high-energy astrophysical processes. For example, active galactic nuclei \cite{27,28}, gamma-ray bursts \cite{29,30}, and ultraluminous X-ray binaries \cite{31} all release enormous amounts of energy. In view of this, in-depth study of the energy extraction mechanisms of such compact objects is of crucial importance for a comprehensive understanding of the astrophysical phenomena they exhibit and their intrinsic nature.

The first energy extraction scheme in history was the Penrose process \cite{12}. This mechanism envisions a particle falling into the black hole from infinity and splitting into two daughter particles within the ergosphere; the one with negative energy falls through the event horizon, while the other, endowed with higher energy than the original particle due to energy conservation, escapes to infinity, thereby extracting the rotational energy of the black hole. Although this theoretical idea is very ingenious, the Penrose process requires that the relative velocity between the two daughter particles exceeds half the speed of light \cite{32}, which is extremely difficult to realize in real astrophysical environments. Therefore, researchers have subsequently proposed a variety of other energy extraction channels \cite{33,34,35,36,37,38,39,83}. In recent years, an energy extraction scheme based on magnetic reconnection, the Comisso-Asenjo mechanism \cite{7}, has attracted widespread attention. Its physical picture can be described as follows: the rapid rotation of the black hole generates antiparallel magnetic fields near the equatorial plane \cite{40,41,42}, and the reversal of the magnetic field direction leads to the formation of current sheets. When the aspect ratio of the current sheet exceeds a critical threshold, it disintegrates due to the plasmoid instability \cite{10,11}. These plasmoids trigger fast magnetic reconnection, converting magnetic energy into the kinetic energy of the plasma and promptly ejecting it from the reconnection region \cite{43,44}. The rotation of the black hole continuously drags the magnetic field lines, thereby inducing new current sheets and allowing the reconnection process to repeat cyclically. In each reconnection event, the plasma within the current sheet is divided into corotating and counter-rotating branches, where the corotating branch is accelerated and the counter-rotating branch is decelerated. Similar to the Penrose process, swallowing the decelerated branch carrying negative energy into the black hole enables the accelerated branch to escape to infinity with the extra energy extracted from the rotational energy of the black hole, thus completing the energy extraction. In some cases, the power of energy extraction of the Comisso-Asenjo mechanism can even surpass that of the Blandford-Znajek mechanism \cite{7}. Currently, this energy extraction mechanism has been extended and applied to various gravitational backgrounds \cite{45,46,47,48,49,50,51,52,53,54,55,56,57,58,59,60,61,62,63,64,65,66,67,68,69,70,71,72,73}.

Although the Comisso-Asenjo mechanism has been extended to so many spacetimes, all the spacetimes involved do not contain the unphysical structure of closed timelike curves. Closed timelike curves, as a typical signature of causality violation in general relativity, were first revealed theoretically by Gödel in his rotating universe model proposed in 1949 \cite{74}. The appearance of such curves generally implies the existence of closed timelike worldlines in the spacetime manifold, thereby allowing particles to return to their own past, which is widely regarded as unphysical causality violation \cite{75}, and is often accompanied by naked singularities. Nevertheless, from the perspective of theoretical exploration, studying the energy extraction process in such spacetimes containing closed timelike curves not only helps to test the universality and robustness of the magnetic reconnection mechanism under extreme and pathological gravitational environments, but also provides a new dynamical criterion for judging the physical plausibility of such metrics.

Although the no-hair theorem \cite{76,77} restricts the uniqueness of Kerr black holes, current observational precision is not yet sufficient to rule out the possibility of non-Kerr compact objects. Alternative configurations such as boson stars \cite{78}, naked singularities, and hairy black holes with anomalous quadrupole moments remain not definitively excluded. Therefore, studying high-energy processes in these non-Kerr backgrounds is of significant astrophysical importance. In particular, the Quevedo-Mashhoon metric \cite{1,2}, as an axisymmetric solution carrying an arbitrary quadrupole moment, provides an ideal theoretical laboratory for testing the performance of the magnetic reconnection mechanism in the non-Kerr case.

Inspired by the above work, in this paper we investigate a class of non-Kerr rotating spacetimes, the Quevedo-Mashhoon spacetime, and conduct an in-depth analysis of the magnetic reconnection process and energy extraction in this spacetime. This metric contains an infinite number of mass and current multipole moments, and possesses a naked singularity and closed timelike curves. Many properties of the Quevedo-Mashhoon spacetime have been extensively studied \cite{4,5,79,80}. In this paper, we aim to study the magnetic reconnection process in this spacetime, explore some unique physical phenomena, and provide theoretical support for the astronomical observation of such black holes.

The rest of this paper is organized as follows. In Section 2, we briefly introduce the Quevedo-Mashhoon spacetime. In Section 3, we present the magnetic reconnection process and energy extraction in this spacetime. In Section 4, we present the power and efficiency of energy extraction. We conclude in Section 5. Throughout this paper, we use natural units $(c=G=1)$.

\section{Introduction to the Quevedo-Mashhoon Spacetime}
The Quevedo-Mashhoon metric is a stationary, axisymmetric, asymptotically flat exact solution of the Einstein field equations \cite{81,82}, which can carry arbitrary mass multipole moments. In its most general form, besides mass and spin, the metric contains an infinite number of additional free parameters that characterize the deviation of its multipole moments from those of the Kerr metric. In this paper, we will focus on a special subclass of the Quevedo-Mashhoon spacetime given in Ref. \cite{3}, which contains only one additional quadrupole moment parameter. In Boyer-Lindquist coordinates, the metric is written as \cite{4,5}
\begin{equation}
ds^{2} = -f\, dt^{2} + \frac{e^{2\gamma}\rho^{2}}{f\Delta} dr^{2} + \frac{e^{2\gamma}\rho^{2}}{f} d\theta^{2} + \left(\frac{\Delta\sin^{2}\theta}{f} - f\omega^{2}\right) d\phi^{2} + 2f\omega\, dt d\phi, 
\end{equation}
where
\begin{equation}
\rho^{2} = (r - M)^{2} - k^{2}\cos^{2}\theta,\quad\Delta = (r - M)^{2} - k^{2}, \quad k = \sqrt{M^2 - a^2},
\end{equation}
\begin{equation}
f = \frac{A}{C} e^{-2\mathbb{Q} P_{2}Q_{2}},\quad \omega = 2a - 2k\frac{D}{A} e^{2\mathbb{Q} P_{2}Q_{2}},\quad e^{2\gamma} = \frac{1}{4}\left(1 + \frac{M}{k}\right)^{2}\frac{A}{x^{2} - 1} e^{2\bar{\gamma}},
\end{equation}
\begin{equation}
A = a_{+}a_{-} + b_{+}b_{-}, \quad C = a_{+}^{2} + b_{+}^{2}, \quad \alpha = \frac{M - k}{a},
\end{equation}
\begin{equation}
\quad D = \alpha x(1 - y^{2})\left(e^{2\mathbb{Q}\delta_{+}} + e^{2\mathbb{Q}\delta_{-}}\right)a_{+} + y(x^{2} - 1)\left[1 - \alpha^{2}e^{2\mathbb{Q}(\delta_{+} + \delta_{-})}\right]b_{+},
\end{equation}
\begin{equation}
\begin{split}
\bar{\gamma} &= \frac{1}{2} (1 + \mathbb{Q})^{2}\ln \frac{x^{2} - 1}{x^{2} - y^{2}}  + 2\mathbb{Q}(1 - P_{2})Q_{1} \\
&\quad+ \mathbb{Q}^{2}(1 - P_{2})\left[(1 + P_{2})(Q_{1}^{2} - Q_{2}^{2})
+ \frac{1}{2} (x^{2} - 1)(2Q_{2}^{2} - 3xQ_{1}Q_{2} + 3Q_{0}Q_{2} - \frac{dQ_2}{dx})\right],
\end{split}
\end{equation}
\begin{equation}
a_{\pm} = x\left[1 - \alpha^{2}e^{2\mathbb{Q}(\delta_{+} + \delta_{-})}\right]\pm \left[1 + \alpha^{2}e^{2\mathbb{Q}(\delta_{+} + \delta_{-})}\right], 
\end{equation}
\begin{equation}
b_{\pm} = \alpha y\left(e^{2\mathbb{Q}\delta_{+}} + e^{2\mathbb{Q}\delta_{-}}\right)\mp \alpha \left(e^{2\mathbb{Q}\delta_{+}} - e^{2\mathbb{Q}\delta_{-}}\right),    
\end{equation}
\begin{equation}
\delta_{\pm} = \frac{1}{2}\ln \frac{(x\pm y)^{2}}{x^{2} - 1} +\frac{3}{2}\left(1 - y^{2}\mp xy\right) + \frac{3}{4}\left[x\left(1 - y^{2}\right)\mp y\left(x^{2} - 1\right)\right]\ln \frac{x - 1}{x + 1},    
\end{equation}
\begin{equation}
x = \frac{r - M}{k},\quad y = \cos \theta .   
\end{equation}
Here, $P_{m}(y)$ and $Q_{m}(x)$ are the Legendre polynomials of the first and second kind of order $m$, respectively. $M$ is the mass of the black hole, $a$ is the spin of the black hole, and $\mathbb{Q}$ is the anomalous quadrupole moment. When $\mathbb{Q} = 0$, the metric reduces to the Kerr metric; when $a = 0$, the metric reduces to the Erez-Rosen metric \cite{6}. When $a=M$, the metric becomes the extreme Kerr spacetime regardless of the value of $\mathbb{Q}$. When $\mathbb{Q} \neq 0$, there is no true event horizon, and a naked singularity appears, which violates the cosmic censorship hypothesis, making it not a standard black hole. Nevertheless, the Quevedo-Mashhoon metric is still suitable for describing the external gravitational field of a rotating object with an arbitrary quadrupole moment, as the naked singularity would not actually appear due to being obscured by matter or quantum effects. The hypothetical event horizon is located at
\begin{equation}
r_h=M+k.   
\end{equation}
The boundary of the ergosphere of the black hole is located at $g_{tt}=0$.

\begin{figure}[!h]
  \centering
  \begin{subfigure}{0.24\textwidth}
    \centering
    \includegraphics[width=\linewidth]{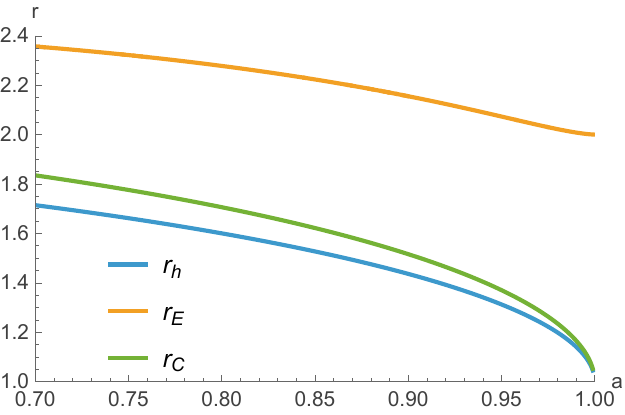}
    \caption{$\mathbb{Q} =-30$}
  \end{subfigure}
  \begin{subfigure}{0.24\textwidth}
    \centering
    \includegraphics[width=\linewidth]{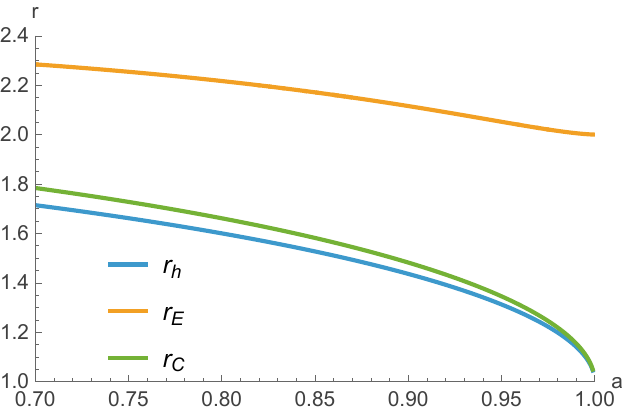}  
     \caption{$\mathbb{Q} =-20$}
  \end{subfigure}
  \begin{subfigure}{0.24\textwidth}
    \centering
    \includegraphics[width=\linewidth]{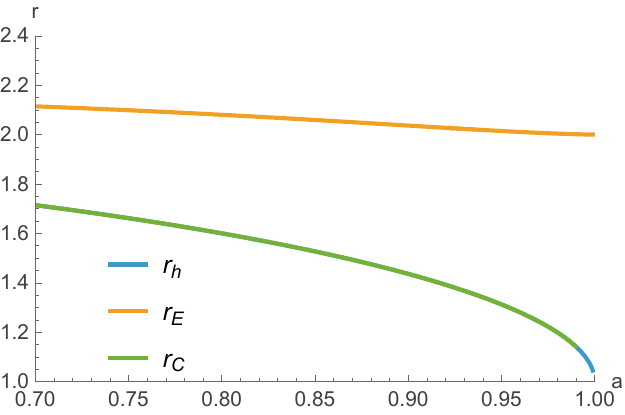}  
     \caption{$\mathbb{Q} =-5$}
  \end{subfigure}
  \begin{subfigure}{0.24\textwidth}
    \centering
    \includegraphics[width=\linewidth]{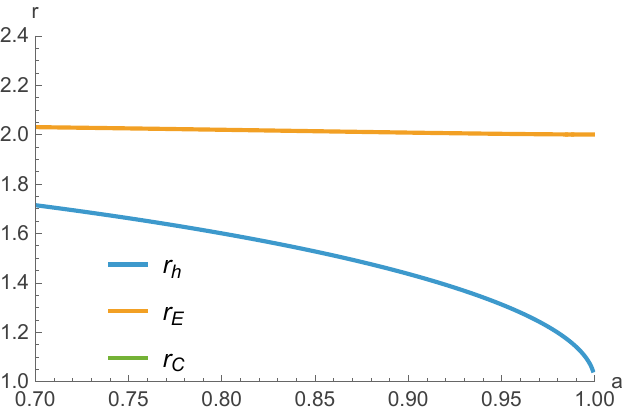}  
     \caption{$\mathbb{Q} =-1$}
  \end{subfigure}
  \begin{subfigure}{0.24\textwidth}
    \centering
    \includegraphics[width=\linewidth]{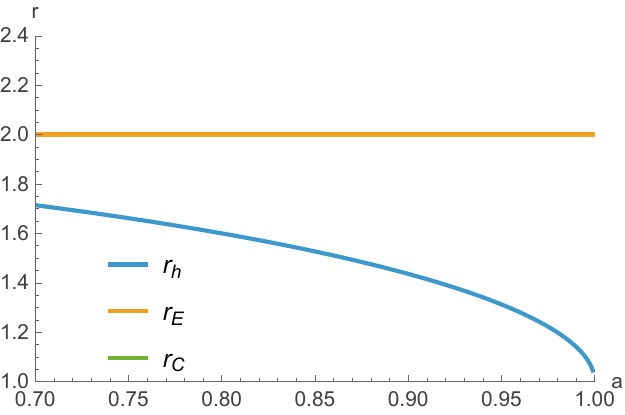}  
     \caption{$\mathbb{Q} =0$(Kerr)}
  \end{subfigure}
\begin{subfigure}{0.24\textwidth}
    \centering
    \includegraphics[width=\linewidth]{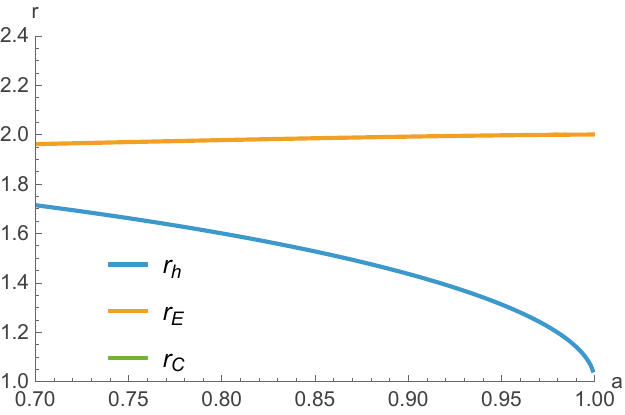}  
     \caption{$\mathbb{Q} =1$}
  \end{subfigure}
\begin{subfigure}{0.24\textwidth}
    \centering
    \includegraphics[width=\linewidth]{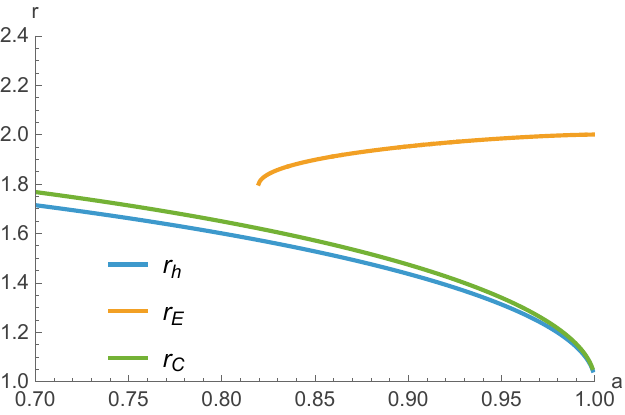}  
     \caption{$\mathbb{Q} =5$}
  \end{subfigure}
\begin{subfigure}{0.24\textwidth}
    \centering
    \includegraphics[width=\linewidth]{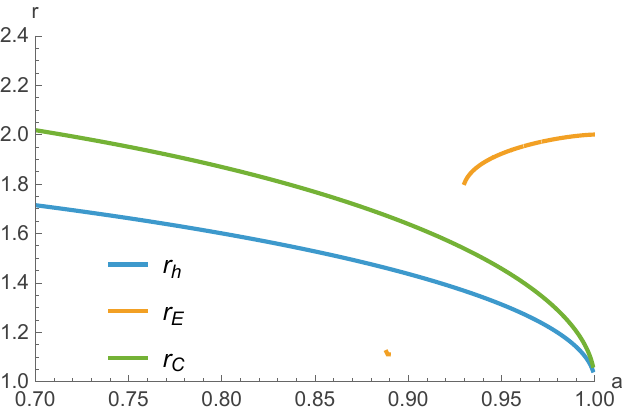} 
     \caption{$\mathbb{Q} =20$}
  \end{subfigure}
\caption{Variation of the event horizon $r_h$, ergosphere boundary $r_E$, and outer radius of the closed timelike curve $r_C$ with spin $a$ for different $\mathbb{Q}$.}
  \label{fig:1}
\end{figure}

On the other hand, unlike many other metrics, the Quevedo-Mashhoon metric may admit a region where $g_{\phi \phi}<0$ outside the hypothetical event horizon, giving rise to closed timelike curves. Closed timelike curves are considered unphysical and must be excluded in calculations. In this paper, without loss of generality, we set $M=1$ and consider the magnetic reconnection process in the equatorial plane, which is precisely the magnetic reconnection process considered by Comisso-Asenjo \cite{7}. In Fig. \ref{fig:1}, we plot the variation of the event horizon, the ergosphere boundary, and the outer radius of the closed timelike curve with spin $a$ in the equatorial plane for different anomalous quadrupole moments.
From Fig. \ref{fig:1}, it can be seen that for $\mathbb{Q}<0$, the larger $|\mathbb{Q}|$ is, the higher the ergosphere; for $\mathbb{Q}>0$, the larger $|\mathbb{Q}|$ is, the lower the ergosphere. For $\mathbb{Q}<0$, the ergosphere always exists, while for $\mathbb{Q}>0$, the ergosphere does not exist for certain spins. When $|\mathbb{Q}|$ is small, closed timelike curves do not exist; when $|\mathbb{Q}|$ is large, closed timelike curves appear, and the larger $|\mathbb{Q}|$ is, the larger the outer radius of the closed timelike curve.

\section{Magnetic Reconnection Process and Energy Extraction in a Non-Kerr Rotating Spacetime with an Anomalous Quadrupole Moment}
In this section, we review the magnetic reconnection process in Ref. \cite{7}, i.e., the Comisso-Asenjo process. We consider a current sheet moving in a circular orbit in the equatorial plane, with the Keplerian angular velocity given by \cite{8}
\begin{equation}
\Omega_K = \frac{- \partial_{r} g_{t\phi} \pm \Bigl[ \bigl( \partial_{r} g_{t\phi} \bigr)^{2} - \partial_{r} g_{tt} \cdot \partial_{r} g_{\phi\phi} \Bigr]^{1/2}}{\partial_{r} g_{\phi\phi}}.
\end{equation}
Here, the positive sign corresponds to prograde orbits, and the negative sign corresponds to retrograde orbits. If the current sheet lies within the ergosphere, only prograde orbits are allowed. In this paper, we consider only prograde orbits. To analyze the energy density of the plasma, it is very convenient to use the zero-angular-momentum-observer (ZAMO) reference frame \cite{9}, which can be written as
\begin{equation}
ds^{2} = \eta_{\mu\nu} \, d\hat{x}^{\mu} d\hat{x}^{\nu}=- d\hat{t}^{2} + \left( d\hat{x}^{1} \right)^{2} + \left( d\hat{x}^{2} \right)^{2} + \left( d\hat{x}^{3} \right)^{2},
\end{equation}
in which
\begin{equation}
d\hat{t} = \kappa dt, \quad d\hat{x}^{i} = -\kappa \beta^{i}  dt + \sqrt{g_{ii}}  dx^{i},
\end{equation}
where
\begin{equation}
\kappa = \sqrt{-g_{tt} + \frac{g_{t\phi}^{2}}{g_{\phi\phi}}}, \quad 
\beta^{\phi} = \frac{\sqrt{g_{\phi\phi}}\,\omega^{\phi}}{\kappa}, \quad 
\omega^{\phi} = -\frac{g_{t\phi}}{g_{\phi\phi}},\quad \beta^{r} =\beta^{\theta} =0.
\end{equation}
Quantities with a hat denote those in the ZAMO frame. Thus, the Keplerian velocity of the current sheet in the ZAMO frame becomes
\begin{equation}
\hat{v}_{K} = \frac{\sqrt{g_{\phi\phi}}}{\kappa}\,\Omega_K - \beta^{\phi}.
\end{equation}
Adopting the single-fluid approximation, the energy-momentum tensor of the fluid can be written as
\begin{equation}
T^{\mu \nu} = p g^{\mu \nu} + \omega_0 U^{\mu} U^{\nu},
\end{equation}
where $p$ and $\omega_0$ represent the proper pressure and enthalpy density of the fluid, and for a relativistically hot plasma, $\omega_0 = 4p$. We have neglected the electromagnetic field tensor part of the fluid, because magnetic reconnection is highly efficient and the energy is completely converted into the kinetic energy of the plasma. $U^{\mu}$ is the four-velocity of the fluid. Using the adiabatic and incompressible fluid approximation, the energy per unit enthalpy at infinity for the accelerated and decelerated plasmas can be expressed as \cite{7}
\begin{equation}
\begin{aligned}
\varepsilon_{\pm} &= \frac{-\kappa g_{\mu 0} T^{\mu 0}}{w_0} \\[2pt]
&= \kappa \hat{\gamma}_{K} \Biggl[ 
    \left(1 + \beta^{\phi} \hat{v}_{K}\right) \sqrt{1+\sigma}
    \pm \cos\xi \left(\beta^{\phi} + \hat{v}_{K}\right) \sqrt{\sigma} - \frac{1}{4\hat{\gamma}_{K}^{2}} \cdot 
    \frac{\sqrt{1+\sigma} \mp \cos\xi \, \hat{v}_{K} \sqrt{\sigma}}
         {1 + \sigma - \cos^{2}\xi \, \hat{v}_{K}^{2} \, \sigma}
\Biggr].
\end{aligned}
\end{equation}
Here, $\xi$ denotes the azimuthal angle of the fluid in the local rest frame, and $\sigma$ is the upstream magnetization parameter of the plasma. $\hat{\gamma}_{K}=\frac{1}{\sqrt{1 - \hat{v}_{K}^{2}}}$ is the Lorentz factor of $\hat{v}_{K}$. From the above expression, it can be seen that $\varepsilon_{\pm}$ is determined by the following parameters: $\{a,\sigma,\xi,r, \mathbb{Q} \}$. Here, $r$ is the radial distance of the dominant $X$-point. According to Ref. \cite{7}, the current sheet in the equatorial plane is susceptible to the plasmoid instability \cite{10,11}, and consequently disintegrates into numerous $X$-point structures. Among these $X$-points, the dominant reconnection $X$-point is located at the intersection of the separatrices that divide the overall reconnection flow. It is this specific $X$-point that governs the dynamical evolution of the reconnection process, i.e., the position corresponding to the radial distance $r$ in the text. Similar to the Penrose process \cite{12}, energy extraction requires that the following two conditions be simultaneously satisfied, namely
\begin{equation}
\varepsilon_{-} < 0 \quad \text{and} \quad \Delta \varepsilon_{+} \equiv \varepsilon_{+} - \left( 1 - \frac{\Gamma}{4(\Gamma-1)} \right) = \varepsilon_{+} > 0.
\end{equation}
where $\Gamma$ is the polytropic index, taken here to be $4/3$.

\begin{figure}[!h]
  \centering
    \includegraphics[width=0.5\linewidth]{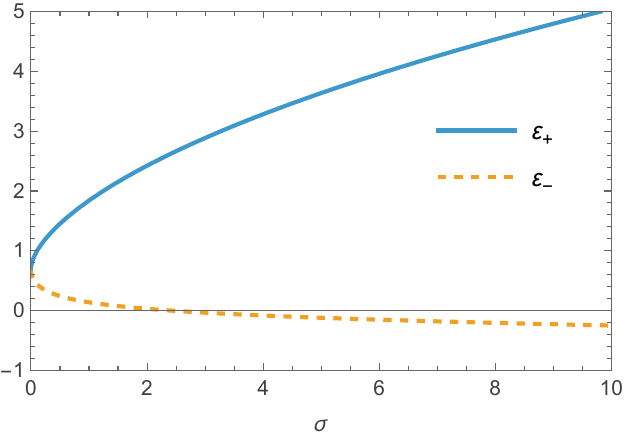} 
  \caption{Variation of $\varepsilon_{\pm}$ with the magnetization parameter. We take $a=0.95, r=1.6>r_s=1.34, \xi=\pi/12, \mathbb{Q}=5$.}
  \label{fig:2}
\end{figure}

\begin{figure}[!h]
  \centering
  \begin{subfigure}{0.45\textwidth}
    \centering
    \includegraphics[width=\linewidth]{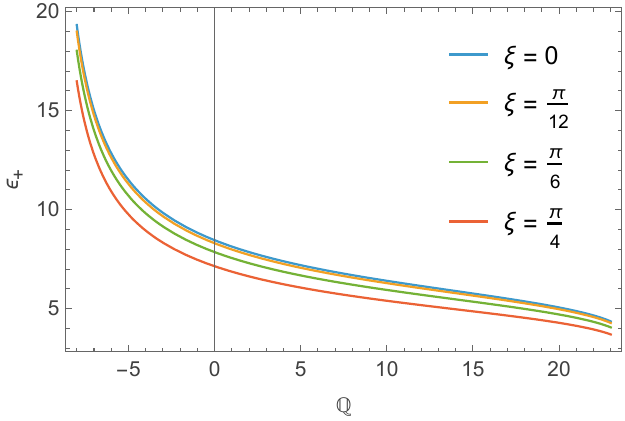}
    \caption{$a=0.95$}
  \end{subfigure}
  \begin{subfigure}{0.45\textwidth}
    \centering
    \includegraphics[width=\linewidth]{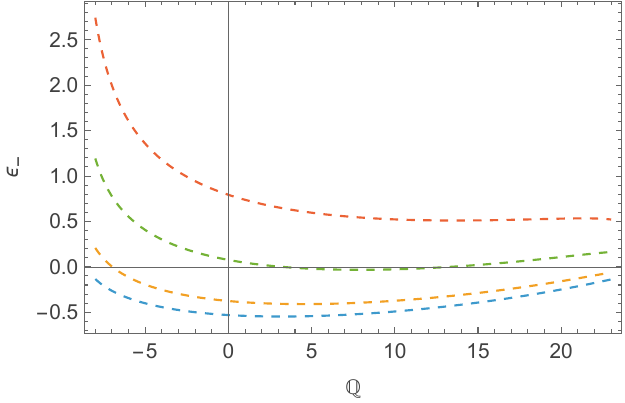}  
     \caption{$a=0.95$}
  \end{subfigure}
  \begin{subfigure}{0.45\textwidth}
    \centering
    \includegraphics[width=\linewidth]{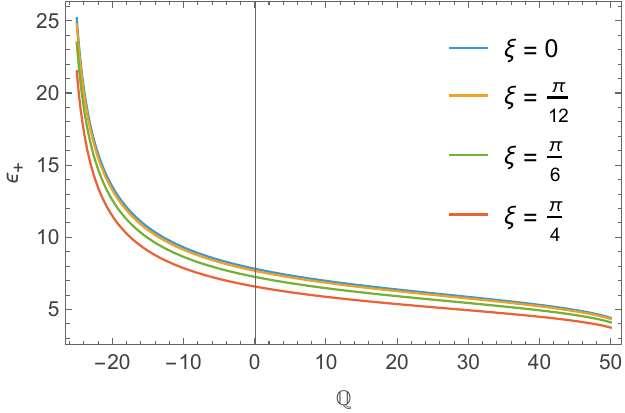}  
     \caption{$a=0.97$}
  \end{subfigure}
  \begin{subfigure}{0.45\textwidth}
    \centering
    \includegraphics[width=\linewidth]{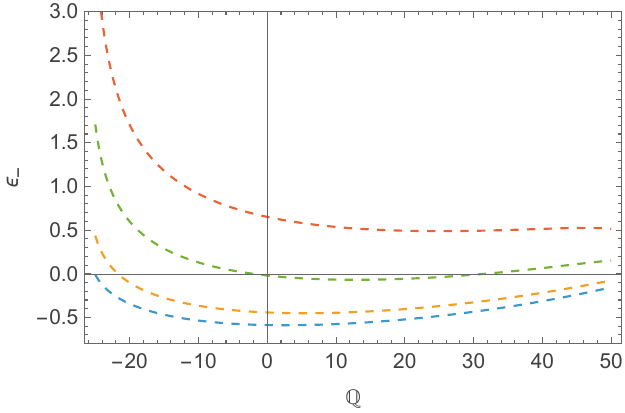}  
     \caption{$a=0.97$}
  \end{subfigure}
\caption{Variation of $\varepsilon_-$ with $\mathbb{Q}$ for different orientation angles, $r=1.6$, $\sigma=20$, top panel: $a=0.95$, bottom panel: $a=0.97$.}
  \label{fig:3}
\end{figure}

Due to the complexity of the metric, the explicit expressions for $\varepsilon_{\pm}$ are very complicated; therefore, we will plot the dependence of $\varepsilon_{\pm}$ on the magnetization parameter in Fig. \ref{fig:2}. Since this metric contains unphysical regions such as closed timelike curves, we adopt the approach used in \cite{13,14} for handling superspinars, and set the radius of the compact object to $r_s = r_c(1+10^{-3})$ to hide these unphysical regions. Therefore, we consider placing the reconnection radius outside the surface of the compact object, i.e., $r \geq r_s$.
From Fig. \ref{fig:2}, it can be seen that $\varepsilon_+$ is always greater than 0, while $\varepsilon_-$ becomes negative only when the magnetization satisfies certain conditions. $\varepsilon_+$ increases with increasing $\sigma$, and $\varepsilon_-$ decreases with increasing $\sigma$. This indicates that larger values of $\sigma$ are more conducive to energy extraction. To analyze the influence of the anomalous quadrupole moment and the magnetic field orientation angle on $\varepsilon_-$, we plot Fig. \ref{fig:3}.
From Fig. \ref{fig:3}, it can be seen that as $\xi$ increases, $\varepsilon_+$ decreases, and $\varepsilon_-$ changes from negative to positive, indicating that a smaller value of $\xi$ is more favorable for energy extraction. Moreover, as $\mathbb{Q}$ gradually increases from negative to positive, $\varepsilon_+$ decreases gradually, while $\varepsilon_-$ first decreases and then increases, with the minimum point biased toward positive $\mathbb{Q}$. This preliminarily suggests that a positive, small anomalous quadrupole moment is more conducive to energy extraction.

Furthermore, we plot the allowed region for energy extraction in the $r-a$ phase space plane, i.e., the region where $\varepsilon_{-} < 0$.
\begin{figure}[!h]
  \centering
  \begin{subfigure}{0.24\textwidth}
    \centering
    \includegraphics[width=\linewidth]{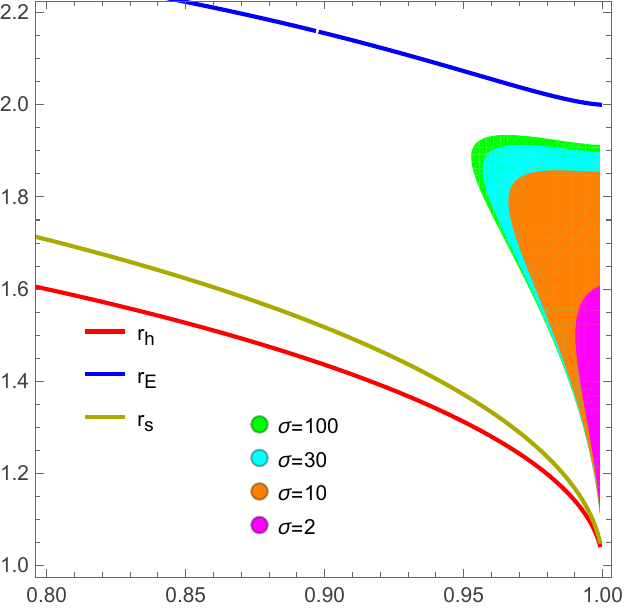}
    \caption{$\mathbb{Q} =-30$}
  \end{subfigure}
  \begin{subfigure}{0.24\textwidth}
    \centering
    \includegraphics[width=\linewidth]{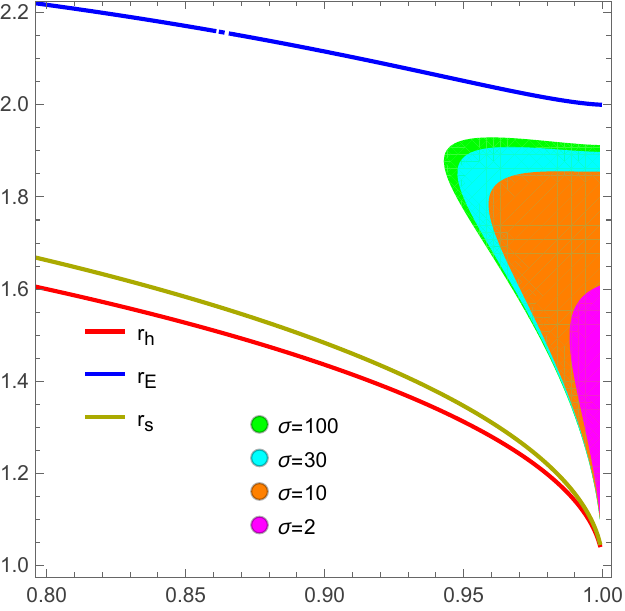}  
     \caption{$\mathbb{Q} =-20$}
  \end{subfigure}
  \begin{subfigure}{0.24\textwidth}
    \centering
    \includegraphics[width=\linewidth]{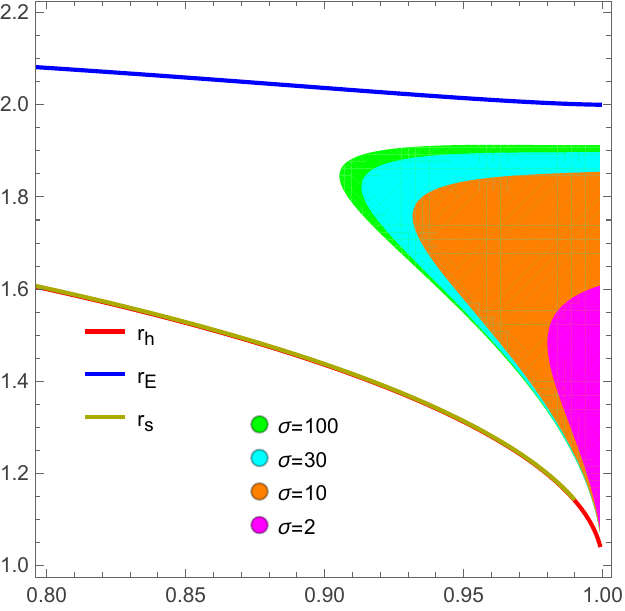}  
     \caption{$\mathbb{Q} =-5$}
  \end{subfigure}
  \begin{subfigure}{0.24\textwidth}
    \centering
    \includegraphics[width=\linewidth]{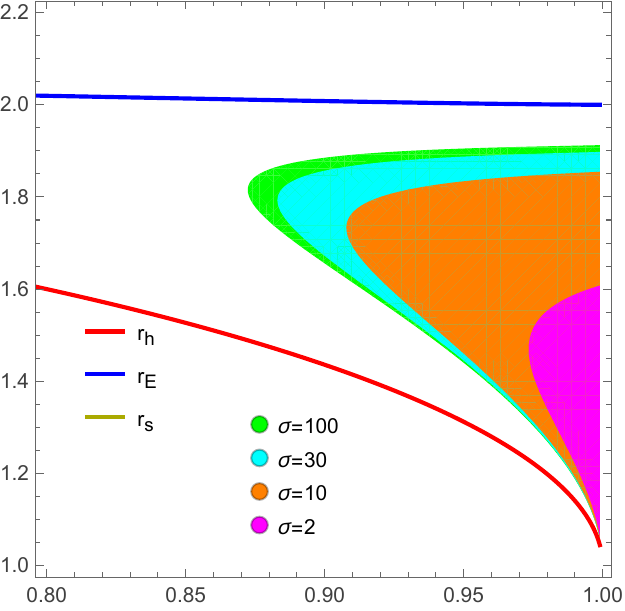}  
     \caption{$\mathbb{Q} =-1$}
  \end{subfigure}
  \begin{subfigure}{0.24\textwidth}
    \centering
    \includegraphics[width=\linewidth]{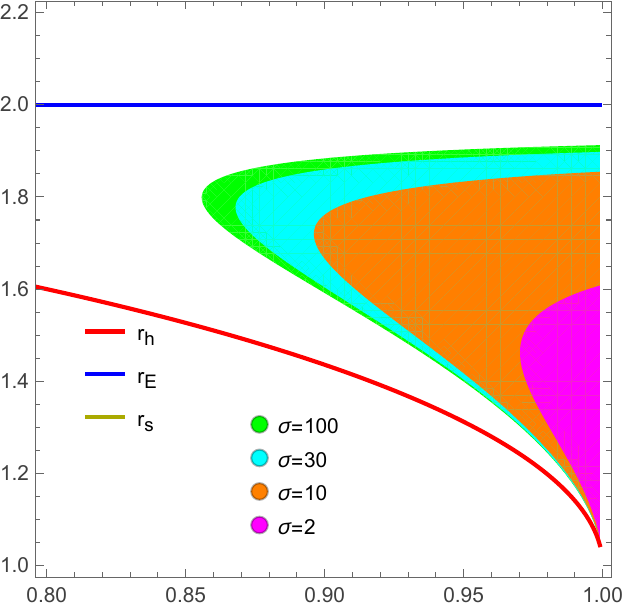}  
     \caption{$\mathbb{Q} =0$(Kerr)}
  \end{subfigure}
\begin{subfigure}{0.24\textwidth}
    \centering
    \includegraphics[width=\linewidth]{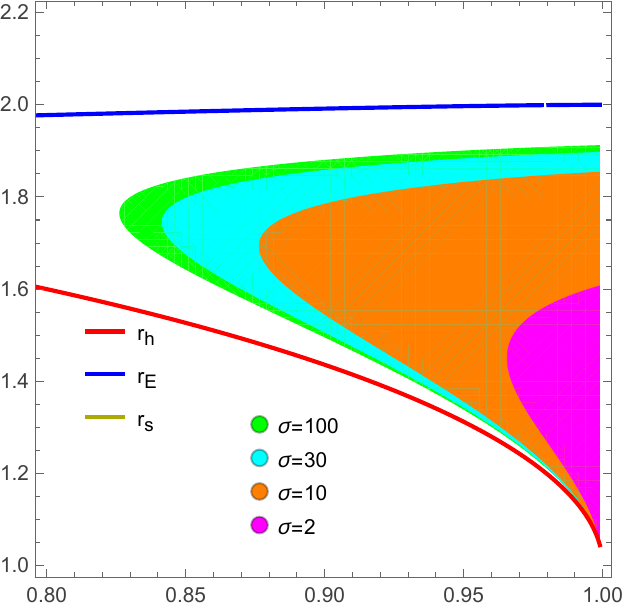}  
     \caption{$\mathbb{Q} =1$}
  \end{subfigure}
\begin{subfigure}{0.24\textwidth}
    \centering
    \includegraphics[width=\linewidth]{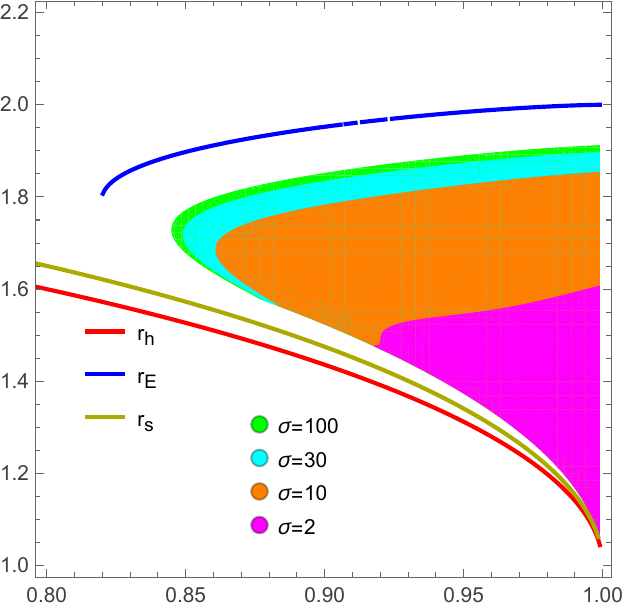}  
     \caption{$\mathbb{Q} =5$}
  \end{subfigure}
\begin{subfigure}{0.24\textwidth}
    \centering
    \includegraphics[width=\linewidth]{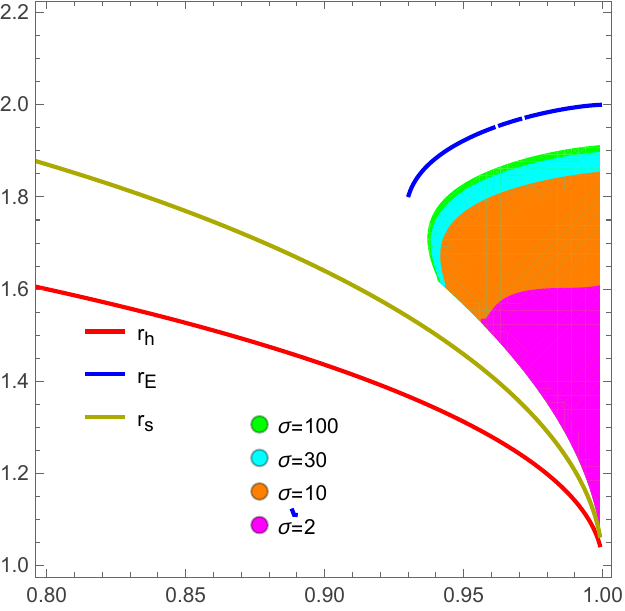} 
     \caption{$\mathbb{Q} =20$}
  \end{subfigure}
\caption{Allowed region for energy extraction in the $r-a$ plane for different $\mathbb{Q}$ and different magnetization parameters $\sigma$, with $\xi=\pi/12$.} 
  \label{fig:4}
\end{figure}
\begin{figure}[!h]
  \centering
  \begin{subfigure}{0.24\textwidth}
    \centering
    \includegraphics[width=\linewidth]{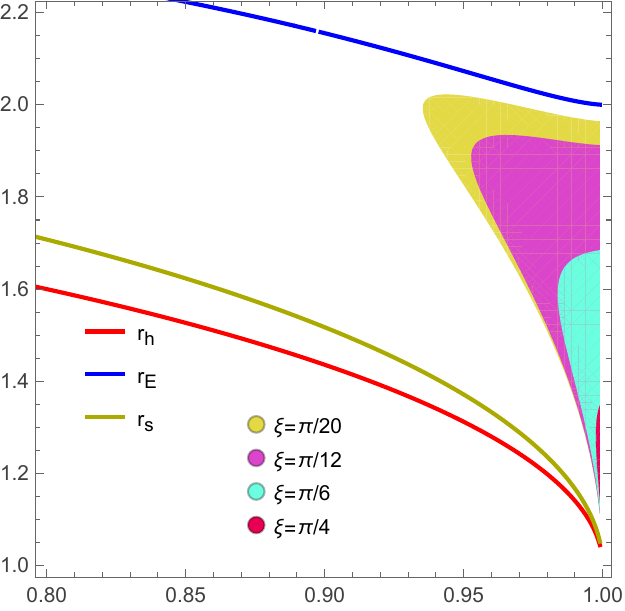}
    \caption{$\mathbb{Q} =-30$}
  \end{subfigure}
  \begin{subfigure}{0.24\textwidth}
    \centering
    \includegraphics[width=\linewidth]{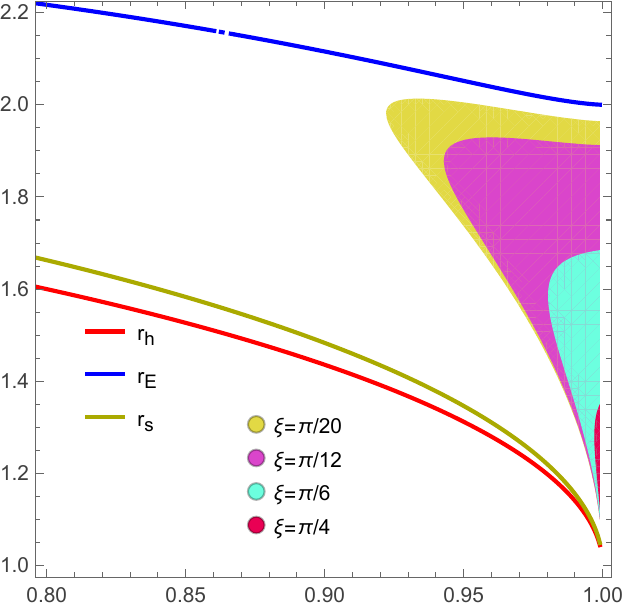}  
     \caption{$\mathbb{Q} =-20$}
  \end{subfigure}
  \begin{subfigure}{0.24\textwidth}
    \centering
    \includegraphics[width=\linewidth]{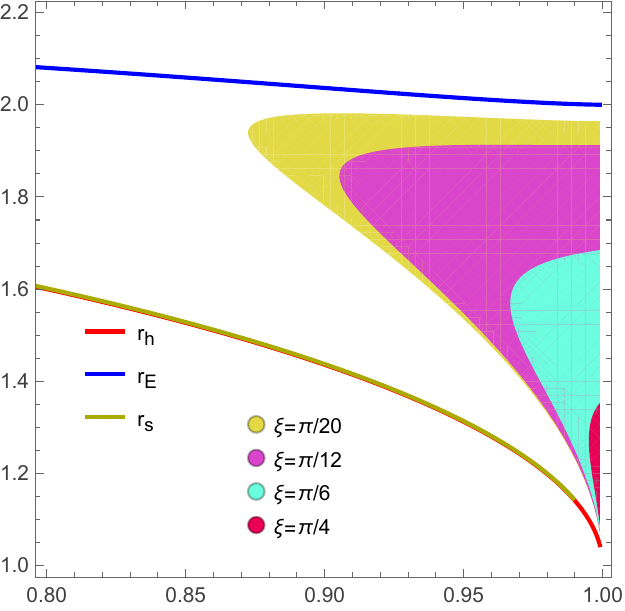}  
     \caption{$\mathbb{Q} =-5$}
  \end{subfigure}
  \begin{subfigure}{0.24\textwidth}
    \centering
    \includegraphics[width=\linewidth]{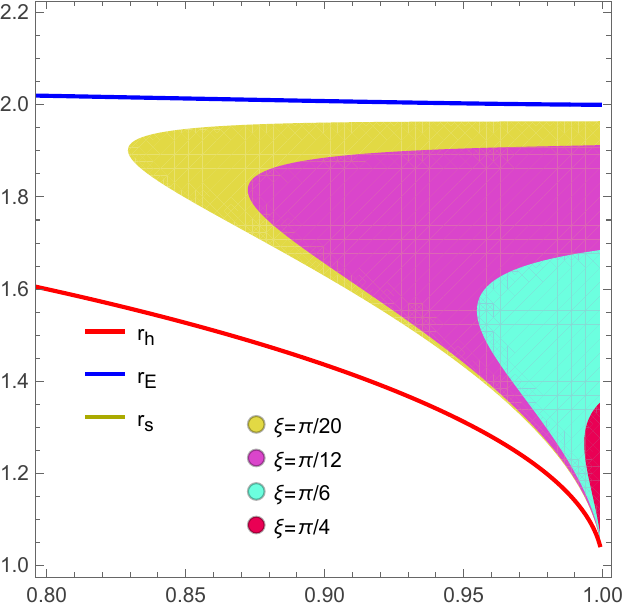}  
     \caption{$\mathbb{Q} =-1$}
  \end{subfigure}
  \begin{subfigure}{0.24\textwidth}
    \centering
    \includegraphics[width=\linewidth]{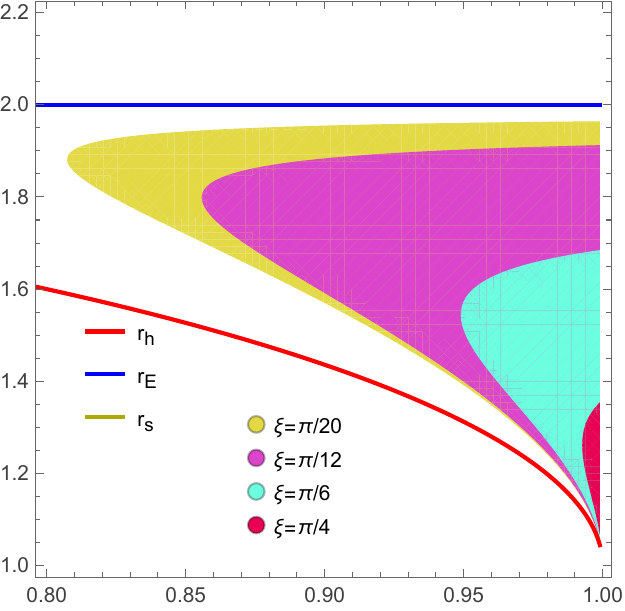}  
     \caption{$\mathbb{Q} =0$(Kerr)}
  \end{subfigure}
\begin{subfigure}{0.24\textwidth}
    \centering
    \includegraphics[width=\linewidth]{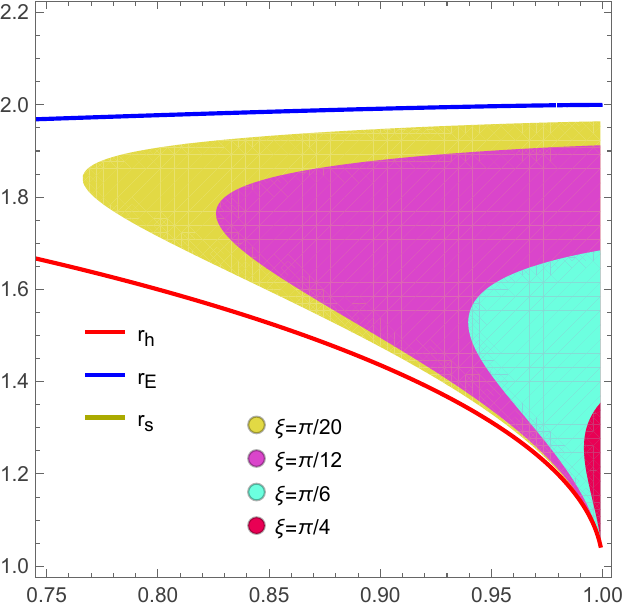}  
     \caption{$\mathbb{Q} =1$}
  \end{subfigure}
\begin{subfigure}{0.24\textwidth}
    \centering
    \includegraphics[width=\linewidth]{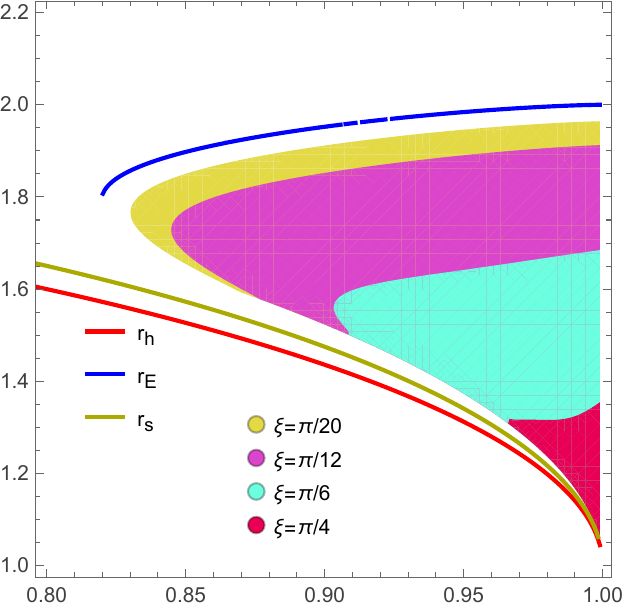}  
     \caption{$\mathbb{Q} =5$}
  \end{subfigure}
\begin{subfigure}{0.24\textwidth}
    \centering
    \includegraphics[width=\linewidth]{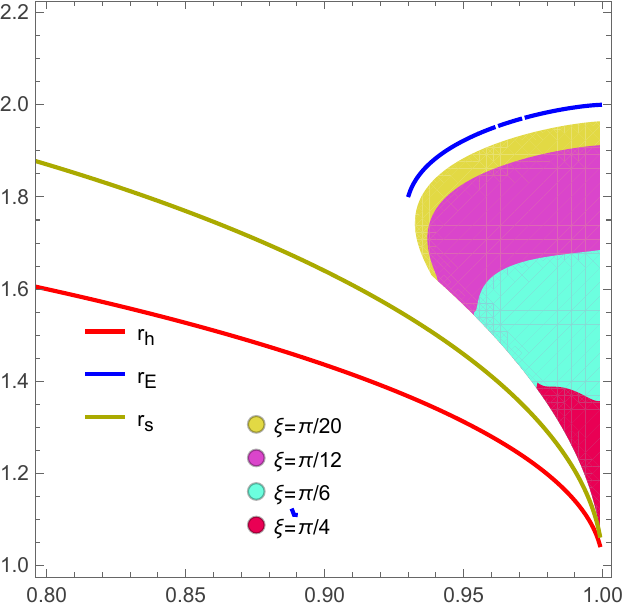} 
     \caption{$\mathbb{Q} =20$}
  \end{subfigure}
\caption{Allowed region for energy extraction in the $r-a$ plane for different $\mathbb{Q}$ and different orientation angles $\xi$, with $\sigma=100$.}
  \label{fig:5}
\end{figure}
From Figs. \ref{fig:4} and \ref{fig:5}, it can be seen that the larger the magnetization parameter or the smaller the magnetic field orientation angle, the larger the allowed region for energy extraction. As $\mathbb{Q}$ gradually changes from negative to 0, the allowed region for energy extraction gradually increases, but remains smaller than that of the Kerr black hole; as $\mathbb{Q}$ gradually increases from 0, the allowed region first increases and then decreases, which again indicates that a positive, small anomalous quadrupole moment is more conducive to energy extraction.

\section{Power and Efficiency of Energy Extraction}
\begin{figure}[!h]
  \centering
  \begin{subfigure}{0.45\textwidth}
    \centering
    \includegraphics[width=\linewidth]{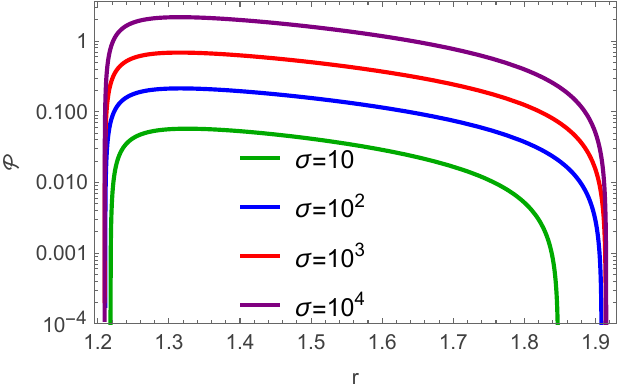}
  \end{subfigure}
  \begin{subfigure}{0.45\textwidth}
    \centering
    \includegraphics[width=\linewidth]{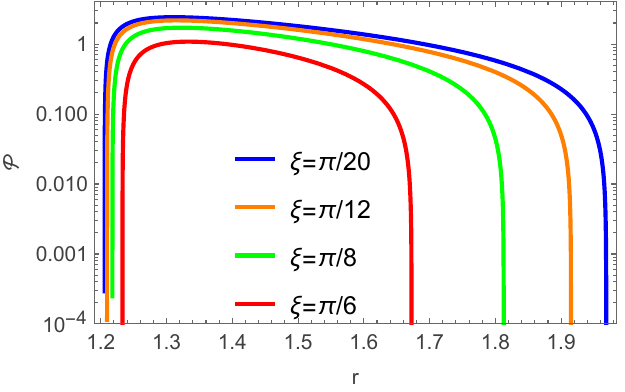}  
  \end{subfigure}
\caption{Left panel: Variation of the energy extraction power with the reconnection radius for different magnetization parameters, $\xi=\pi/12$. Right panel: Variation of the energy extraction power with the reconnection radius for different magnetic field orientation angles, $\sigma=10000$. Common parameters: $a=0.98, \mathbb{Q}=1$.}
  \label{fig:6}
\end{figure}

In this section, we present the power and efficiency of energy extraction for this black hole. The power per unit enthalpy density of energy extraction is given by \cite{7}
\begin{equation}
\mathcal{P}=-\varepsilon_-A_{in}U_{in}.
\end{equation}
For the collisionless regime, $U_{in}\approx 0.1$ \cite{15}, while for the collisional regime, $U_{in}\approx 0.01$ \cite{16}. In this paper, we consider the collisionless case. $A_{in}$ is the cross-sectional area of the inflowing plasma, which can be estimated as $A_{in}\approx r_E^2-r_s^2$. Fig. \ref{fig:6} shows the variation of the energy extraction power with the reconnection radius for different magnetization parameters and magnetic field orientation angles. It can be seen that a larger magnetization parameter and a smaller magnetic field orientation angle lead to higher power, and the power strongly depends on the reconnection radius.
\begin{figure}[!h]
  \centering
  \begin{subfigure}{0.45\textwidth}
    \centering
    \includegraphics[width=\linewidth]{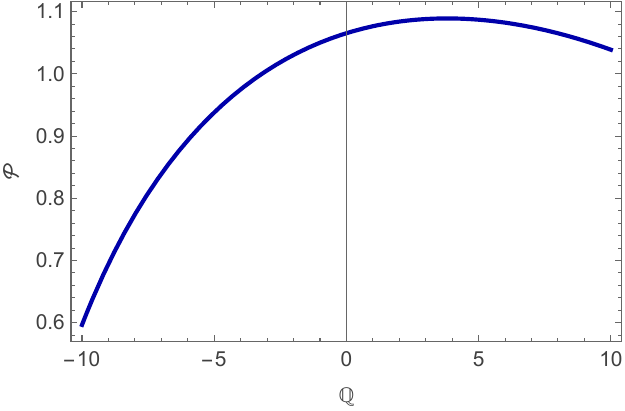}
  \end{subfigure}
  \begin{subfigure}{0.45\textwidth}
    \centering
    \includegraphics[width=\linewidth]{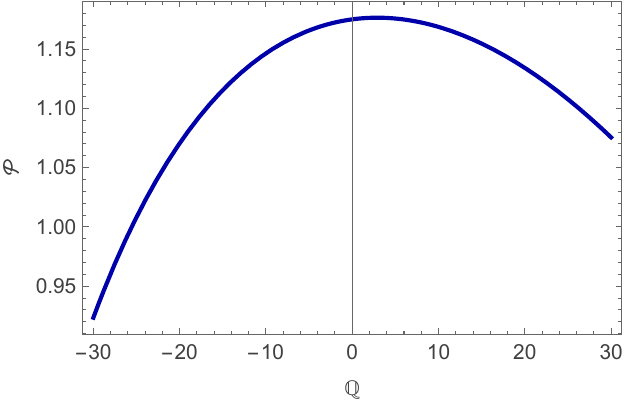}  
  \end{subfigure}
\caption{Variation of the energy extraction power with the anomalous quadrupole moment, left panel: $a=0.96$, right panel: $a=0.98$. Common parameters: $r=1.6, \xi=\pi/12, \sigma=10000$.}
  \label{fig:7}
\end{figure}
In Fig. \ref{fig:7}, we also plot the variation of the energy extraction power with $\mathbb{Q}$. The curve exhibits a downward-opening parabola, indicating that the power of energy extraction first increases and then decreases, but the maximum is located to the right of zero, which once again shows that a positive, small anomalous quadrupole moment is conducive to energy extraction. Moreover, the power at $a=0.98$ is higher than that at $a=0.96$. This is because the higher the spin, the higher the power.

Next, we present the energy extraction efficiency. The efficiency of energy extraction is given by \cite{7}
\begin{equation}
\eta=\frac{\varepsilon_+}{\varepsilon_++\varepsilon_-}.
\end{equation}
From the above equation, it can be seen that for the energy extraction efficiency to be meaningful, $\eta$ must be greater than 1.
\begin{figure}[!h]
  \centering
  \begin{subfigure}{0.45\textwidth}
    \centering
    \includegraphics[width=\linewidth]{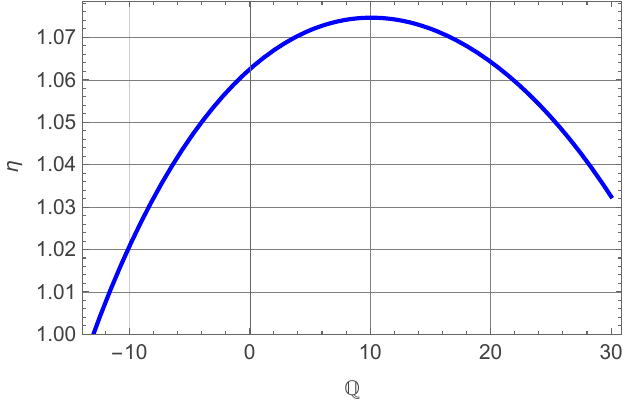}
  \end{subfigure}
  \begin{subfigure}{0.45\textwidth}
    \centering
    \includegraphics[width=\linewidth]{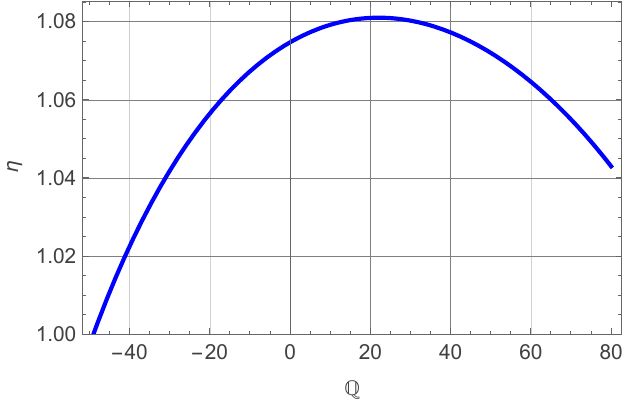}  
  \end{subfigure}
\caption{Variation of the energy extraction efficiency with the anomalous quadrupole moment, left panel: $a=0.96$, right panel: $a=0.98$. Common parameters: $r=1.6, \xi=\pi/12, \sigma=100$.}
  \label{fig:8}
\end{figure}
From Fig. \ref{fig:8}, it can be seen that the variation of efficiency with $\mathbb{Q}$ is similar to the variation of power with $\mathbb{Q}$, both first increasing and then decreasing. When $a=0.96$, the efficiency for $\mathbb{Q}$ in the range $0-20.6$ is higher than that of the Kerr black hole, and when $a=0.98$, the efficiency for $\mathbb{Q}$ in the range $0-45.5$ is higher than that of the Kerr black hole. These conclusions also indicate that a positive and small anomalous quadrupole moment possesses a stronger energy extraction capability, which is consistent with the main theme of this paper. Moreover, a higher spin leads to a higher efficiency.

\section{Conclusion}
In this paper, we have investigated energy extraction via the Comisso-Asenjo magnetic reconnection mechanism in the Quevedo-Mashhoon spacetime, which is accompanied by an anomalous quadrupole moment. Although the Comisso-Asenjo mechanism has been extended to many spacetimes, we perform energy extraction for the first time in a spacetime possessing a naked singularity and closed timelike curves. Although the naked singularity and closed timelike curves are unphysical, the metric can still be used to describe the gravitational field outside a compact object.

We first plotted the hypothetical event horizon, the closed timelike curve, and the ergosphere boundary in the equatorial plane of this spacetime, thereby gaining a preliminary understanding of the spacetime characteristics and energy extraction. After reviewing the magnetic reconnection mechanism, we analyzed the energy per unit enthalpy at infinity for the accelerated and decelerated plasmas, the allowed region for energy extraction, and the power and efficiency of energy extraction, all of which are profoundly influenced by the spin, magnetization, magnetic field orientation angle, reconnection radius, and anomalous quadrupole moment. Among them, this spacetime shares some common features with other spacetimes, for example, a higher spin, a larger magnetization parameter, a smaller magnetic field orientation angle, and a moderate reconnection radius are more favorable for energy extraction. Although energy extraction is possible for both positive and negative anomalous quadrupole moments, a positive and small anomalous quadrupole moment corresponds to a stronger energy extraction capability. This is because, for a negative anomalous quadrupole moment, the energy per unit enthalpy at infinity of the decelerated plasma, the allowed region for energy extraction, and the power and efficiency of energy extraction are all lower than the corresponding parts for a Kerr black hole, and the more negative it is, the lower they are compared to those of a Kerr black hole; however, for a positive anomalous quadrupole moment, the energy per unit enthalpy at infinity of the decelerated plasma, the allowed region for energy extraction, and the power and efficiency of energy extraction are higher than those of a Kerr black hole for a small anomalous quadrupole moment, but lower than those of a Kerr black hole for a large anomalous quadrupole moment.

Currently, Refs. \cite{17,18,19,20,21} combine the magnetic reconnection mechanism with hot spot imaging to identify signals of energy extraction from astronomical observations. Ref. \cite{22} demonstrated that under realistic astrophysical conditions, the magnetic reconnection-driven Penrose process is energetically feasible, thus supporting the view that it may occur. We look forward to further progress in future studies on the magnetic reconnection-driven Penrose process.

\noindent {\bf Acknowledgments}

\noindent
This work is supported by the National Natural Science Foundation of China (Grants Nos. 12375043, 12575069 ), and Chongqing Normal University Fund Project (Grants No. 26XLB001).

\end{document}